# Composite Fermion States around 2D Hole Landau Level Filling Factor 3/2 in Tilted Magnetic Fields

**Po Zhang**

*International Center for Quantum Materials, School of Physics, Peking University, Beijing 100871, China*

**Ruiyuan Liu**

*International Center for Quantum Materials, School of Physics, Peking University, Beijing 100871, China*

*Department of Physics and Astronomy, Rice University, Houston, Texas 77251-1892, USA*

**Rui-Rui Du**

*Department of Physics and Astronomy, Rice University, Houston, Texas 77251-1892, USA*

*International Center for Quantum Materials, School of Physics, Peking University, Beijing 100871, China*

*Collaborative Innovation Center of Quantum Matter, Beijing 100871, China*

**L. N. Pfeiffer** and **K. W. West**

*Department of Electrical Engineering, Princeton University, Princeton, New Jersey 08544, USA*

## Abstract

Transport measurements under a tilted magnetic field were performed on a series of C-doped (001) AlGaAs/GaAs/AlGaAs two-dimensional hole samples. Due to a large g-factor, Zeeman energy is large and comparable to the cyclotron energy in these samples. On the other hand, it was found that the in-plane component $g_{//}$ is small, and the effect of tilted magnetic field is mainly to increase the effective mass of holes. We investigate the spin transition of composite fermion states around Landau level (LL) filling factor 3/2. We found that the ν = 4/3 state encounters a partial to full spin polarization transition, conforming to the same pattern as that of electron samples. In addition, high-resistance phase emerges at ν = 3/2 under very high tilt angles. We interpret both of these phenomena as a consequence of LL crossings that are mainly driven by the orbital effects. The roles that the spin degrees of freedom play in FQH states around ν = 3/2 in these systems will be discussed.

The composite fermion (CF) model[1,2] has made a great success in descripting the fractional quantum Hall effect (FQHE), especially for the p/(2p ± 1) series around the half-filling of the lowest LL. In this model, a CF is obtained by mathematically attaching 2 magnetic flux quanta to an electron. The attached fluxes, so called Chern-Simons fluxes, cancel the applied magnetic field and make the effective field felt by CFs zero at corresponding even-denominator filling factor $v = 1/2$. The problem of FQHE then reduces to CF's integer quantum hall effect. Experiments have confirmed CFs with an effective mass[3,4], a well-defined Fermi vector, and consequently, semi-classic motion[5,6]. Angular dependent magneto-transport measurement provides a way to investigate the LL spectrum of CFs (so called Λ-Level, ΛL). Around $v = 3/2$ in GaAs two-dimensional electron gas (2DEG) it has been revealed by this method that CF has a g-factor largely the same as that of the underlying electrons, and ν and 2 − ν FQH states are counterparts of particle-hole symmetry with the same filling factor of ΛL[7]. However, it remains an interesting question whether these spin-related properties of CFs still hold for systems with larger g-factors, especially ones with Zeeman energy comparable to or even larger than cyclotron energy.

In GaAs materials, two dimensional hole gas (2DHG) has a larger g-factor than 2DEG. The bulk value of g-factor for electrons and heavy holes are $g = 0.44$ and $\kappa = 1.2$ respectively[8–10]. The predicted $g_\perp$ value for 2DHG confined in GaAs is $6\kappa = 7.2$[11,12]. However the measured value of g is usually smaller than 7.2[13–15]. There are several works investigating CF states around $v = 3/2$ on Si-doped (113) GaAs 2DHG reporting very similar behavior with those in 2DEG[16–18]. The g-factor taken for data analyzes is not so large (1.1 ~ 1.2) in these works. 2DHG in C-doped (001) GaAs provides another alternative. It is reported that the effective g-factor in this class of 2DHG is large enough to cause LL crossing around 1T, with the magnetic field solely perpendicular to the sample plane[19]. (001) GaAs substrate also provides higher symmetry than (113), hence a more isotropic in-plane g-factor[12]. Here we perform angular dependent magneto-transport measurements on a series of MBE grown, C-doped, p-type (001) GaAs quantum well samples around $v = 3/2$.

Four samples, denoted by A-D and covering a density range of 0.7 ~ 1.5 (in units of $10^{11} cm^{-2}$ throughout this paper) were measured in this experiment. Table I lists relevant information for each sample. All devices are Van der Pauw squares, each with eight In/Zn contacts diffused symmetrically on the perimeter. The specimen was immersed into the coolants of a dilution refrigerator with a base temperature T ~ 20mK. The specimen was mounted on a rotator, which facilitated the sample *in situ* rotation around an axis perpendicular to the magnetic field. Standard four-terminal lock-in technique was used in the experiments, with a measurement current of 10 nA. An overview of transport traces from sample C and the experimental configuration is depicted in Fig. 1. The tilt angle θ is defined as the angle between magnetic field and normal of the 2DHG plane.

States around $v = 3/2$ under tilted magnetic field for sample C and D is shown in Fig. 2. The two samples showed significant difference during the tilting. As illustrated, $R_{xx}$ minimum and $R_{xy}$ plateau of 4/3 state in sample C disappear first and reappear at larger θ while for sample D, whose carrier density is higher, the traces show no significant change until an insulating phase (IP) emerges.

A detailed plot of $R_{xx}$ at fixed filling factor versus $B_{tot}$ for different samples is shown in Fig. 3a. For samples A-C whose densities p < 1, $R_{xx}(4/3)$ shows a peak that exceeds the value of $R_{xx}(3/2)$, indicating a partial- to full- spin polarization transition. $R_{xx}(5/3)$ keeps unchanged or initially drops due to the gap enhancement by an increasing

Zeeman energy, showing no signatures for spin transition. $R_{XX}(4/3)$ always exceeds $R_{XX}(5/3)$ after its transition, suggesting a change in their gaps' relative magnitude. However, for sample D, $R_{XX}$ of 4/3 is always larger than that of 5/3 throughout the region until an IP appears. This case for D suggests that even at the zero tilt the transition has already happened. Another evidence the transition has already happened is that there is no peak of $R_{XX}(4/3)$ before the IP comes up in sample D.

One may recall previous works on 2DEG[7] and 2DHG in (311) GaAs[16–18], in which almost the same evolution of 4/3 and 5/3 states was observed. However, due to the fact that the Zeeman energy in C-doped 2DHG is so large, it exceeds cyclotron energy even at zero tilt in these samples. It should be instructive to look at relevant energy scales here. GaAs 2DEG has a g-factor of 0.44, effective mass of 0.067, leading to a Zeeman energy of 0.3*K/T* and cyclotron energy of 20.1*K/T*. For holes in Ref. 16 hosted by a Si-doped (311) GaAs, the authors took values $g = 1.1$ and $m* = 0.38$, so $E_z = 0.73K/T$, $\hbar\omega_c = 3.5K/T$. Zeeman energy is small compared with cyclotron energy in both systems. However, for holes in C-doped (100) GaAs QWs, Ref. 19 reported a g-factor between 5 ~ 7.2 and $m* = 0.4$, making Zeeman energy (3.4 ~ 4.8 K/T) larger than LL gap (3.4 K/T) at least in the region $B \approx 1T$.

It should be noted that in Ref. 19 the authors only analyze the LLs around $B_\perp = 1T$ with a filling factor of about 9, a density $p = 2.2$. The LL spectrum in 2DHG is non-linear and complicated. Even though the Zeeman energy is large, there is still a distance to getting the conclusion that, in C-doped (100) GaAs 2DHG systems, FQH states around 3/2 take place in $|N=1,\uparrow\rangle$ Landau level rather than usual $|N=0,\downarrow\rangle$ level, as illustrated in Fig. 3(b). However, this is the most likely picture. One can define an effective Zeeman energy $E_z^* = |E_z - \hbar\omega_c|$ that characterizes the spin-splitting gap between the LL where 4/3 state stays and its nearest neighbor. If Zeeman energy is slightly larger than LL gap, $E_z^*$ will increase with $B_{tot}$ at fixed filling factor, inducing the spin transition of 4/3 state. A Zeeman energy slightly smaller than LL gap will also induce a spin transition of 4/3 state for a similar reason, but $E_z^*$ decreases at first. However, the dropping of $R_{xx}(5/3)$ at the beginning of tilting and the fact that $R_{xx}(5/3)$ goes smaller than $R_{xx}(4/3)$ after the latter's transition (Fig. 3(a)) require an increasing $E_z^*$ with increasing $B_{tot}$, since 5/3 is a spin-polarized state (there is no transition of 5/3 observed) and a larger $E_z^*$ helps enlarging the gap. The stripe phase we found at higher tilted field also supports a Zeeman energy larger than LL gap, which we will discuss later.

If we take the lowest LL in Fig. 3(b) away (the right-bottom level), an analogue will arise between state $v$ in our system and state $v-1$ in small g system. It is interesting that these analogue states $v$ and $v-1$ have different ΛL filling factors. We denote filling factor of CFs by $V$, $V$ s of 4/3 and 5/3 states are 2 and 1 in our samples, inferred by the number of transition times. However, $V$ s of 4/3-1 and 5/3-1, i.e. 1/3 and 2/3 state, are reversed (1 and 2) in small g system, as in 2DEG 1/3 is polarized state while 2/3 is partially polarized. It looks like the particle-hole symmetry in small g system that makes $v$ and $2-v$ being analogues with the same $V$ still holds for our samples. However, since the LL where 3/2 state stays in our samples has a different index (N=1) and interacts with a different LL ($|N=0,\downarrow\rangle$), there is little reason that the '$2-v$ symmetry' should still survive. A more reasonable alternate is that the reversed ΛL spectrum arises from interacting of $|N=0,\downarrow\rangle$ and $|N=1,\uparrow\rangle$.

The relationship between $1/\cos\theta_c$ and carrier density p (also $B_{4/3}$, the out-of-plane magnetic field where $v$=4/3) is illustrated in Fig. 3(c), $\theta_c$ stands for the critical $\theta$ where 4/3 state disappears. We find a linear relationship between $1/\cos\theta_c$ and p. It is worthwhile to compare this linear relationship with that in an ideal 2D system. For an ideal 2D system with small g-factor, the gap of ΛL for 4/3 state is proportional to Coulomb energy, that is $\propto e^2/4\pi\varepsilon_0\varepsilon_r l_B \propto \sqrt{B_\perp} \propto \sqrt{p}$. And the Zeeman energy is $g\mu_B B_{tot} \propto p/\cos\theta$. The transition of 4/3 happens when these two energies are equal, leading to $\cos\theta_c \propto \sqrt{p}$, which is very different from our case. If we consider an ideal 2D system with large g-factor that its Zeeman energy is slightly larger than gap of LLs, we get $\frac{1}{\cos\theta_c} = \frac{5}{g}(14.80\alpha\frac{1}{\sqrt{B}}+1)$ by just replacing $E_z$ with $E_z^*$, here $\alpha$ is the ratio of ΛL gap to Coulomb energy, approx. 0.024 in n-type GaAs heterojunctions[20], and we take $\epsilon_r = \epsilon_{GaAs} = 13.1$. The relationship is also plotted in Fig. 3(c) (dashed line for g=5.1 and dotted for g=7), assuming that the curve passes through the point $(B_{4/3}, 1/\cos\theta_c) = (3.04, 1)$. This relation is also very different from the linear curve, confirming that Zeeman energy is non-linear with $B_{tot}$, which has been reported on 2DHG. The linear relationship of $1/\cos\theta_c$ vs p may arise from the non-linear change of effective mass in holes, which requires more experimental data to make things clear.

We now turn to the IP phase. Fig. 4 illustrates the insulating phase at high tilting angles. The IP was most pronounced in sample D, so we took a detail measurement of it on this sample. Fig. 5(a) shows anisotropy in the order of 1000:1 for the IP, suggesting a stripe phase induced by in-plane magnetic field. In our knowledge, however, this kind of stripe phase in GaAs 2D system has only been found at $v = 5/2$ and $v = 7/2$ before[21], i.e. in N=1 LL. We interpret it as a consequence of LL level crossing. Fig. 3(b) shows a sketch of the crossing for this condition. When $|N=1,\uparrow\rangle$ LL goes below $|N=0,\downarrow\rangle$ LL, the strip phase at $v=3/2$ is then hosted by N=1 LL, so an in-plane field induced IP may occur. Unlike in electron system where such IPs appear in pair, another 'strip phase' hosted by N=1 LL (at $v = 9/2$) was not found in our samples (see Fig. 5(b)), maybe due to the screening from three LLs bellow it or just the sample quality. In Ref. 22 the authors reported an observation of strip phase without in-plane field at $v=7/2$ while the usual one at $v = 9/2$ is missing. They explained it as a consequence of spin-orbit coupling. However, this can also be explained by our simplistic model that LLs hosting 3/2 and 5/2 (also 7/2 and 9/2) are swapped by the LL crossing. 7/2 state in 2DHG behaves similar to 9/2 in 2DEG, so the IP appears at $v=7/2$.

The stripe phase also provides an evidence that the crossing has already happened in zero tilt. If not, since the LLs are reversed at high fields for the IPs to occur, there must be an angle where the LLs just cross. $E_z^*$ will decrease first and then increase, so does the gap of spin-polarized state 5/3. However, $R_{xx}$ of 5/3 only decreases at the beginning of tilting, does not increase at all. The crossing of LL will also have influence on gap of all other FQH states around 3/2, which has not been observed, neither.

In conclusion, by performing angular dependent magneto-transport measurements on four C-doped p-type (100) GaAs quantum well samples, we observed spin transition of 4/3 state and found an in-plane magnetic field induced stripe phase at $v = 3/2$. They can both be understood in a scenario of crossed LL in zero tilt magnetic fields.

The work at Peking University was supported by National Basic Research Program of China (No. 2012CB921301 and 2014CB920901). The work at Rice was supported by Welch Foundation Grant (No. C-1682). The work at Princeton was partially funded by the Gordon and Betty Moore Foundation as well as the National Science Foundation MRSEC Program through the Princeton Center for Complex Materials (Grant No. DMR-0819860). A portion of this work was performed at the National High Magnetic Field Laboratory, which is supported by National Science Foundation Cooperative Agreement No. DMR-1157490 and the State of Florida.

# References

[1] J. K. Jain, Phys. Rev. Lett. **63**, 199 (1989).

[2] B. I. Halperin, P. A. Lee, and N. Read, Phys. Rev. B **47**, 7312 (1993).

[3] R. R. Du, H. L. Stormer, D. C. Tsui, L. N. Pfeiffer, and K. W. West, Phys. Rev. Lett. **70**, 2944 (1993).

[4] H. C. Manoharan, M. Shayegan, and S. J. Klepper, Phys. Rev. Lett. **73**, 3270 (1994).

[5] R. L. Willett, R. R. Ruel, K. W. West, and L. N. Pfeiffer, Phys. Rev. Lett. **71**, 3846 (1993).

[6] W. Kang, H. L. Stormer, L. N. Pfeiffer, K. W. Baldwin, and K. W. West, Phys. Rev. Lett. **71**, 3850 (1993).

[7] R. R. Du, A. S. Yeh, H. L. Stormer, D. C. Tsui, L. N. Pfeiffer, and K. W. West, Phys. Rev. Lett. **75**, 3926 (1995).

[8]

[9] W. Ekardt, K. Lösch, and D. Bimberg, Phys. Rev. B **20**, 3303 (1979).

[10] M. J. Snelling, G. P. Flinn, A. S. Plaut, R. T. Harley, A. C. Tropper, R. Eccleston, and C. C. Phillips, Phys. Rev. B **44**, 11345 (1991).

[11] R. T. Harley, "Coherent optical interactions in semiconductors," (Springer US, Boston, MA, 1994) Chap. Spin-Related Effects in III-V Semiconductors, pp. 91–109.

[12] J. M. Luttinger, Phys. Rev. **102**, 1030 (1956).

[13] R. Winkler, S. J. Papadakis, E. P. De Poortere, and M. Shayegan, Phys. Rev. Lett. **85**, 4574 (2000).

[14] S. Glasberg, H. Shtrikman, and I. Bar-Joseph, Phys. Rev. B **63**, 201308 (2001).

[15] H. W. van Kesteren, E. C. Cosman, W. A. J. A. van der Poel, and C. T. Foxon, Phys. Rev. B **41**, 5283 (1990).

[16] V. F. Sapega, M. Cardona, K. Ploog, E. L. Ivchenko, and D. N. Mirlin, Phys. Rev. B **45**, 4320 (1992).

[17] K. Muraki and Y. Hirayama, Phys. Rev. B **59**, R2502 (1999).

[18] P. J. Rodgers, B. L. Gallagher, M. Henini, and G. Hill, Journal of Physics: Condensed Matter **5**, L565 (1993).

[19] A. G. Davies, R. Newbury, M. Pepper, J. E. F. Frost, D. A. Ritchie, and G. A. C. Jones, Phys. Rev. B **44**, 13128 (1991).

[20] Z. Q. Yuan, R. R. Du, M. J. Manfra, L. N. Pfeiffer, and K. W. West, Applied Physics Letters **94**, 052103 (2009), http://dx.doi.org/10.1063/1.3077147.

[21] Y. Liu, S. Hasdemir, A. W´ojs, J. K. Jain, L. N. Pfeiffer, K. W. West, K. W. Baldwin, and M. Shayegan, Phys. Rev. B **90**, 085301 (2014).

[22] W. Pan, R. R. Du, H. L. Stormer, D. C. Tsui, L. N. Pfeiffer, K. W. Baldwin, and K. W. West, Phys. Rev. Lett. **83**, 820 (1999).

[23] M. J. Manfra, R. de Picciotto, Z. Jiang, S. H. Simon, L. N. Pfeiffer, K. W. West, and A. M. Sergent, Phys. Rev. Lett. **98**, 206804 (2007).

TABLE I. Information of four samples. All of them are cleaved from MBE grown, C-doped, p-type (100) GaAs quantum well wafers. B and C are cleaved from the same wafer.

|  | A | B | C | D |
|---|---|---|---|---|
| width of QW (nm) | 17.5 | 17.5 | 17.5 | 20 |
| density ($10^{11}$ cm$^{-2}$) | 0.70 | 0.86 | 0.94 | 1.54 |
| mobility ($10^6$ cm$^2$/Vs) | 0.6 | 1.1 | 1.3 | 1.7 |
| doping[a] | a | s | s | s |

[a] a for one-side asymmetric doping, s for two-sides symmetric doping

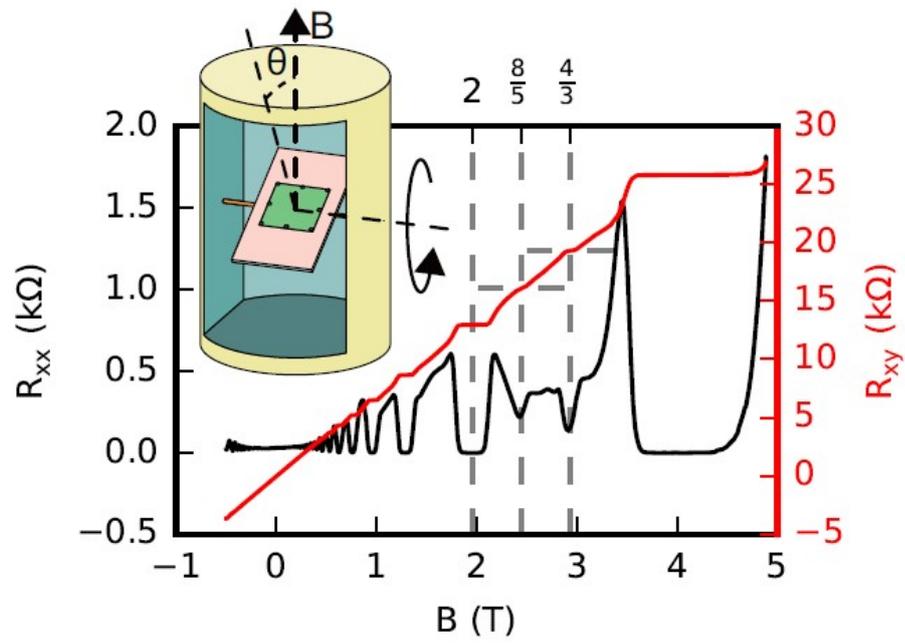

FIG. 1. An overview of $R_{xx}$ and $R_{xy}$ at zero tilt in sample C. A schematic diagram of experimental configuration is shown in the inset. The Van der Pauw sample was mounted on a rotatable stage and immersed in the coolants of a dilution refrigerator.

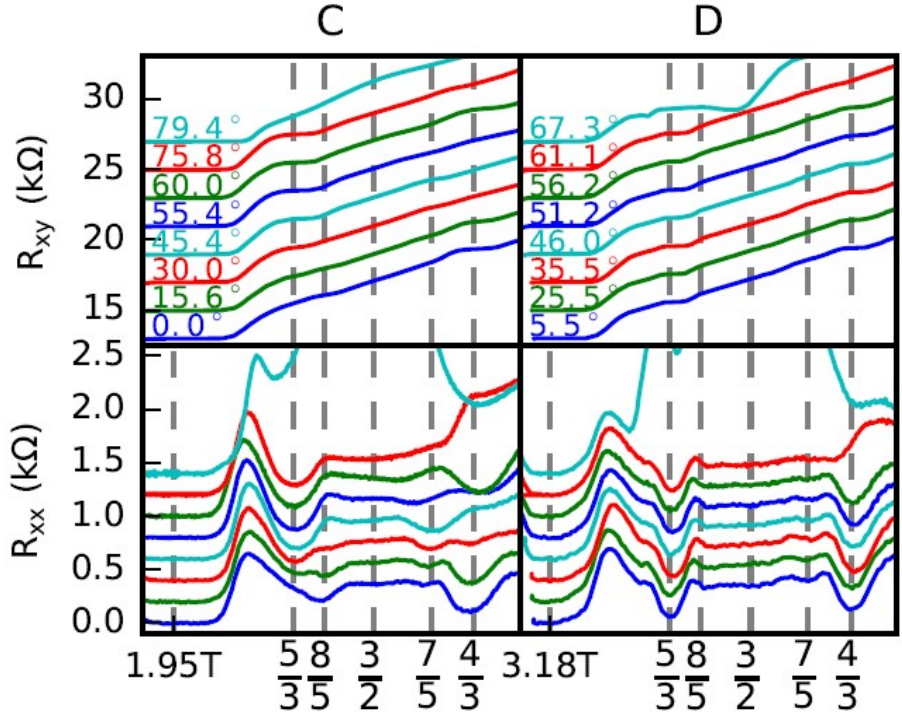

FIG. 2. An overview of $R_{xx}$ and $R_{xy}$ under tilted magnetic field for sample C and D. Sample C shows a transition of 4/3 state as well as a gap enlargement of 5/3 state while sample D shows no significant changes in these states until at higher angles insulating phases arise around $v = 3/2$ in both samples. Curves are shifted vertically for clarify. The values of tilt angle $\theta$ are noted above $R_{xy}$ curves. The labels on lateral axis are the $B_\perp$ value of $v = 2$ followed by the filling factors.

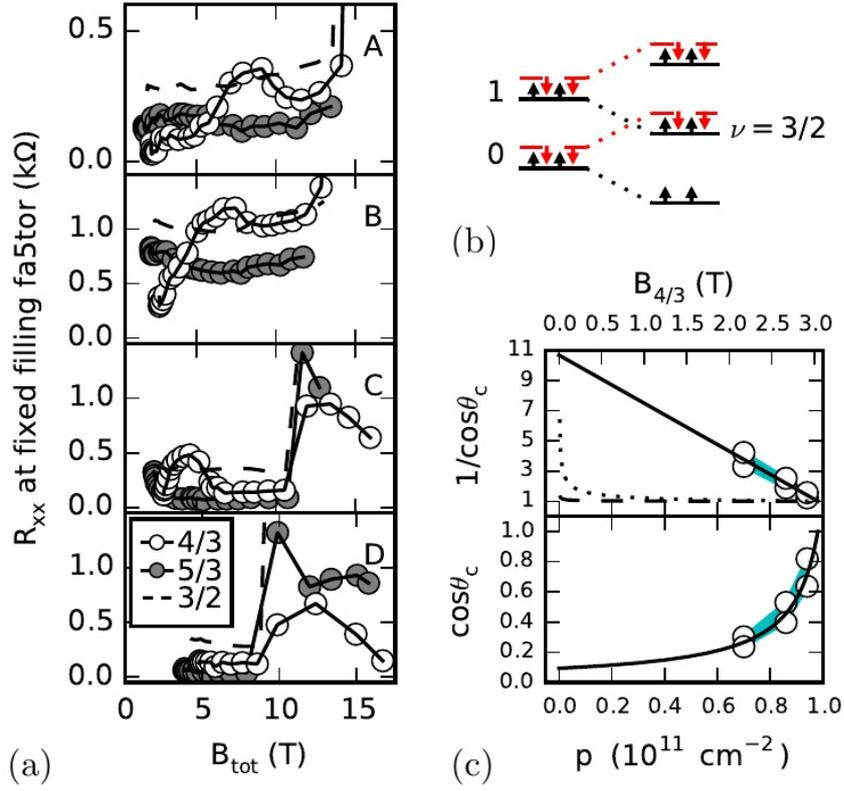

FIG. 3. (a) $R_{xx}$ at fixed filling factor versus $B_{tot}$ for different samples. Legend is depicted in the bottom panel. (b) A sketch of LL spectrum in systems with small g-factors ( left ), and large g-factors (right) like ours. (c) $1/\cos\theta_c$ versus density p (and $B_\perp$ at $\nu = 4/3$) shows a very linear relationship. $\theta_c$ is the critical angle where 4/3 transition takes place. 2 points are used to determine $\theta_c$ for each sample. One is the smallest $\theta$ where minimum of 4/3 state disappears in our $R_{xx}$ vs B curve and the other is the largest. The dashed and dotted lines are calculated from a simple model with linear Zeeman energy which has been discussed in the text.

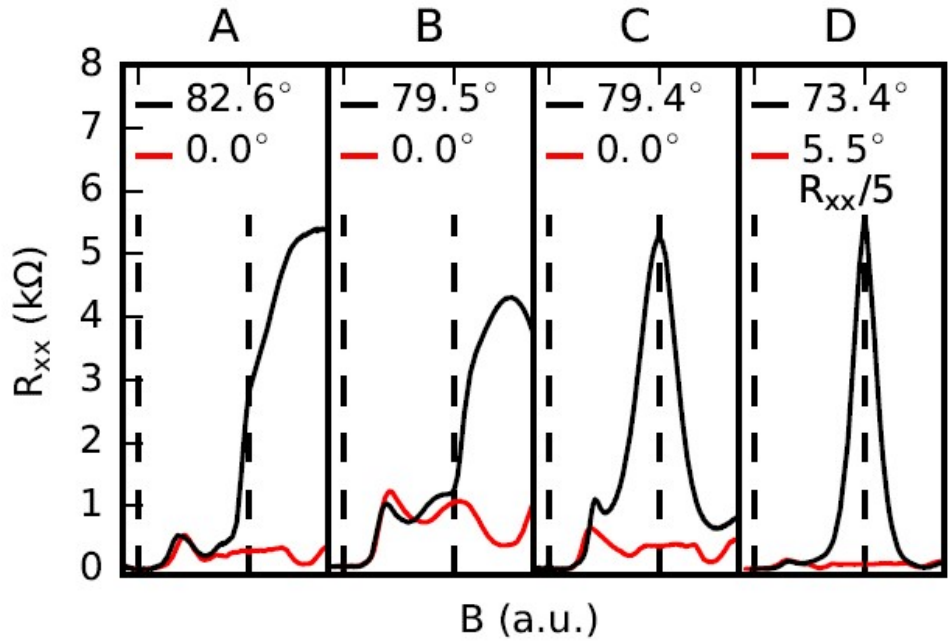

FIG. 4. An overview of insulating phase. The insulating phase comes up at high tilting angles in all four samples, around $v = 3/2$. $R_{xx}$ for sample D is divided by 5, here the 'x' in the subscript means that the current is parallel to the in-plane magnetic field. From left to right, the dashed vertical lines in each subfigure indicate positions where $v = 2$ (left) and $3/2$ (right).

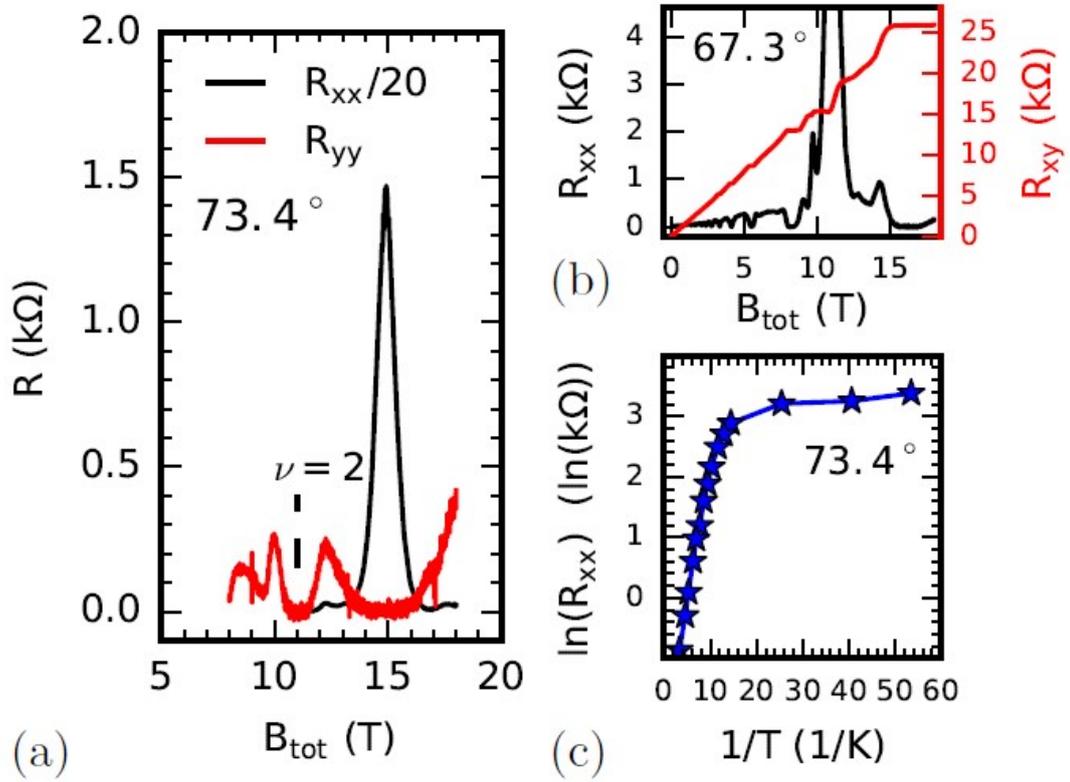

FIG. 5. Details of insulating phase (IP) in sample D. (a) Anisotropy between $R_{xx}$ and $R_{yy}$, indicates a stripe phase. 'x' in the subscript means that the current is parallel to the in-plane magnetic field while 'y' means perpendicular. (b) Only a single IP appears in the region $v > 1$. (c) Temperature dependence of IP. The slope in linear region is 0.44 K. Tilting angles are noted on each subfigure.